\documentclass[graybox,natbib,nosecnum]{svmult}
\bibpunct{(}{)}{;}{a}{}{,} 

\pdfoutput=1

\usepackage{mathptmx}       
\usepackage{helvet}         
\usepackage{courier}        
\usepackage{type1cm}

\usepackage{makeidx}        
\usepackage{graphicx}       
                            
\usepackage{multicol}       
\usepackage[bottom]{footmisc}
\usepackage[normalem]{ulem} 
\usepackage{hyperref}

\usepackage{soul}
\usepackage{rotating}

\begin{document}
\title*{Exoplanet Atmospheres at High Spectral Resolution}
\author{Jayne L. Birkby}
\institute{Jayne L. Birkby \at Assistant Professor \at Anton Pannekoek Insitute of Astronomy, University of Amsterdam, Science Park 904, Amsterdam, 1098XH, The Netherlands, \email{jbirkby@uva.nl}}

\maketitle

\abstract{The spectrum of an exoplanet reveals the physical, chemical,
  and biological processes that have shaped its history and govern its
  future. However, observations of exoplanet spectra are complicated
  by the overwhelming glare of their host stars. This review chapter
  focuses on high resolution spectroscopy (HRS) ($R=25,000-100,000$),
  which helps to disentangle and isolate the exoplanet's spectrum. At
  high spectral resolution, molecular features are resolved into a
  dense forest of individual lines in a pattern that is unique for a
  given molecule. For close-in planets, the spectral lines undergo
  large Doppler-shifts during the planet's orbit, while the host star
  and Earth's spectral features remain essentially stationary,
  enabling a velocity separation of the planet. For slower-moving,
  wide-orbit planets, HRS, aided by high contrast imaging, instead
  isolates their spectra using their spatial separation. The lines in
  the exoplanet spectrum are detected by comparing them with high
  resolution spectra from atmospheric modelling codes; essentially a
  form of fingerprinting for exoplanet atmospheres. This measures the
  planet's orbital velocity, and helps define its true mass and
  orbital inclination. Consequently, HRS can detect both transiting
  and non-transiting planets. It also simultaneously characterizes the
  planet's atmosphere due to its sensitivity to the depth, shape, and
  position of the planet's spectral lines. These are altered by the
  planet's atmospheric composition, structure, clouds, and dynamics,
  including day-to-night winds and its rotation period. This chapter
  describes the HRS technique in detail, highlighting its successes in
  exoplanet detection and characterization, and concludes with the
  future prospects of using HRS to identify biomarkers on nearby rocky
  worlds, and map features in the atmospheres of giant exoplanets.}

\section{Introduction}
\subsection{Exploring the exoplanet zoo}
It is extremely challenging to directly observe the spectrum of an
exoplanet. Not only are exoplanets faint objects, but they are
typically located at sub-arcsecond orbital separations, buried in the
glare of their host stars that outshine them by factors of a few
thousand to billions. However, the spectrum of an exoplanet is the key
to understanding the physical, chemical, and biological processes that
have shaped its history and govern its future. Therefore,
spectroscopic detection is enormously valuable in our quest to
understand the striking diversity of the exoplanet zoo
(e.g. \citealt{Ful17,VanE17}), and in our search for life beyond the
Solar system. There are novel ways in which we can infer an
exoplanet's spectrum based on how it affects the total light of the
star-planet system throughout its orbit. Although these approaches
have delivered the majority of the observed exoplanet spectra to date
(see e.g. \citealt{Sin16,Cross17}), they rely crucially on the planet
being favourably aligned so that it passes in front or behind its host
star along our line-of-sight, i.e. it forms a transit or secondary
eclipse (see Kreidberg's chapter of this handbook for more discussion
on this technique). The chance of this occurring in nature depends
predominantly on the stellar radius ($R_{\star}$) and the orbital
separation ($a$) of the star-planet system:
$P_{transit}\sim R_{\star}/a$. For any planets orbiting the $\sim350$
main-sequence stars in our local 10 pc neighbourhood ($\sim33$ ly;
RECONS, \citealt{Hen16,Hen18}), this chance is typically small (less
than 2 percent). Statistical analyses predict that the nearest
transiting, potentially habitable, Earth-size planet is located at
$\sim11\,$pc \citep{Dress15}, which makes the seven Earth-size planets
orbiting the ultracool dwarf star TRAPPIST-1 at $12.1\pm0.4\,$pc
particularly interesting \citep{Gil17}. However, the nearest
\emph{non-transiting}, potentially habitable Earth-size planets, are
predicted to be an order of magnitude closer to us, making the
non-transiting planetary systems of Proxima Centauri and Ross 128 at
just $1.3\,$pc and $3.4\,$pc respectively, prime targets in our search
for life \citep{Ang16,Bon17}. Therefore, in this chapter, we focus on
alternative approaches that can disentangle and isolate the light
emitted or reflected by the planet itself, regardless of its orbital
orientation to our line-of-sight, allowing us to explore the exoplanet
zoo and our local neighbourhood in full.
 
Traditionally, the isolation of planetary light from its host star has
been achieved by physically blocking out or nulling the host star
using a coronagraph. This has allowed the imaging of widely separated
exoplanets (sep$>0.1^{\prime\prime}$, $a>1$AU) in broadband filters
and more recently spectroscopy at low-to-moderate spectral resolution
($R=\lambda/\Delta\lambda\sim<4000$; see
e.g. \citealt{Kono13,Maci15}). Biller \& Bonnefoy discuss this in
detail in their chapter of this handbook. Here, we will instead
investigate how we can use moderate-to-\emph{high-resolution
  spectroscopy} ($R=5000-100,000$) to study light not only from
widely-separated exoplanets, but also those on very close-in orbits
(sep$<0.005^{\prime\prime}$, $a<0.05\,$AU). This is important for our
nearest neighbours orbiting the $\sim250$ cool M-dwarf stars within
$10\,$pc, whose habitable zones are much closer to the host star in
comparison to the Sun, with a year lasting on the order of a few tens
of days. The techniques we will discuss in this chapter enable not
only the spectroscopic detection of the planet and measurement of its
true mass, but also a detailed study of its atmospheric properties,
including its composition, structure, and its dynamics, including
global wind patterns and its rotation period.

\section{Exoplanets at high resolution}
\subsection{Close-in exoplanets: the Doppler Dance}
\begin{figure}
\centering
\includegraphics[width=\textwidth]{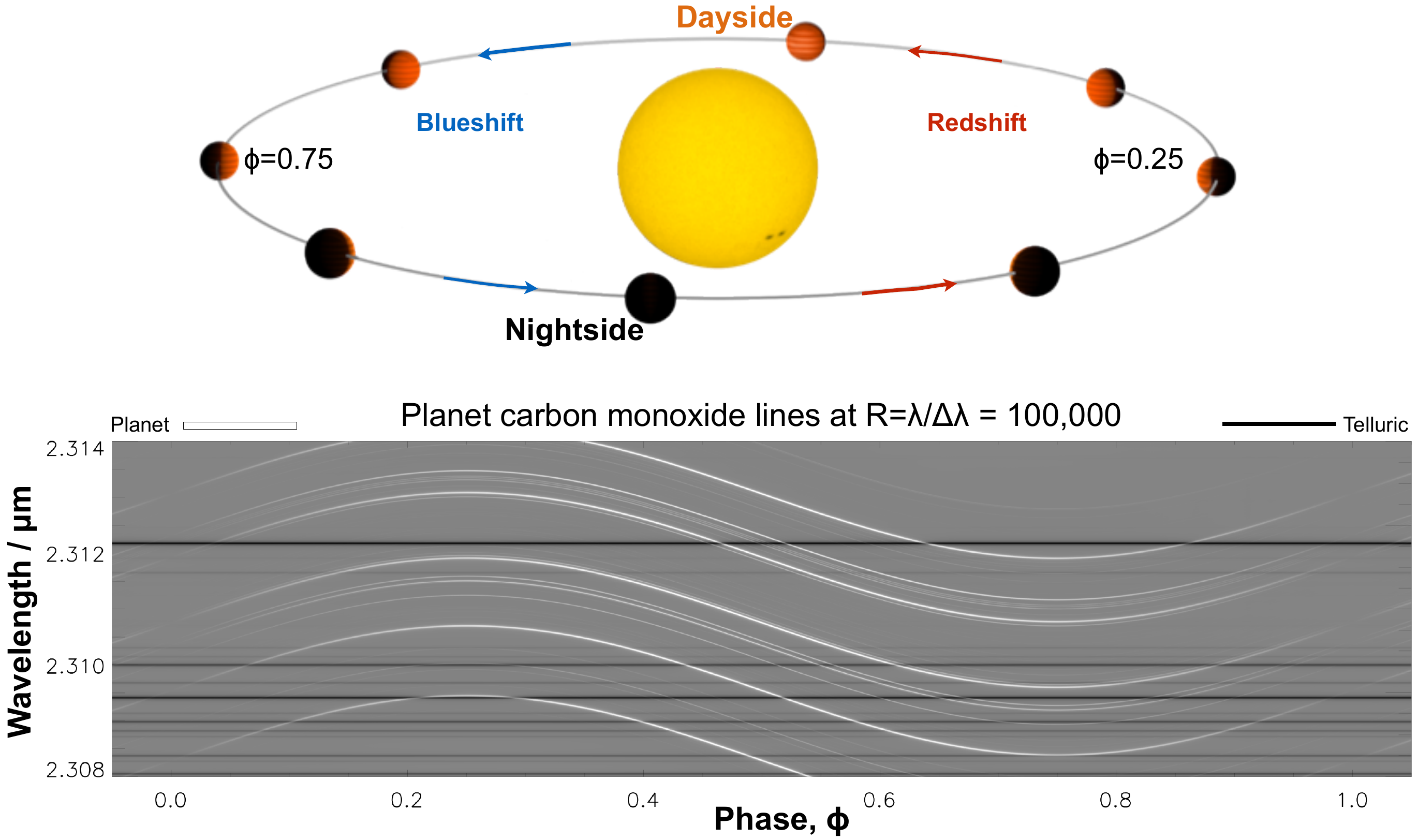}
\caption{The high-resolution spectroscopy (HRS) technique. The top
  panel shows the illumination of a non-transiting planet throughout
  its orbit, and highlights when its spectrum is red- or
  blue-shifting. At high-resolution, the planet's spectrum is resolved
  into a dense forest of individual lines in a pattern that is unique
  to each molecule (see Figure~\ref{fig:resolutions}). These lines
  trace out the radial velocity curve of the planet
  ($\Delta$RV$\sim$km/s), allowing it to be robustly disentangled from
  the essentially stationary stellar lines ($\Delta$RV$\sim$m/s) and
  the contaminating spectral lines from Earth's telluric features. The
  maximum rate of change of velocity is reached around superior and
  inferior conjunction. At the quadrature points highlighted in the
  top panel, the spectral lines of the close-in planet are maximally
  separated in velocity from the host star and tellurics, but the
  change in the Doppler shift of the planet lines is small here making
  it difficult to disentangle them from the tellurics. The planet can
  be observed via its thermally emitted spectrum, or by the spectrum
  it reflects from the host star. Star-planet image \copyright Ernst
  de Mooij.}
\label{fig:technique1}
\end{figure}

The high-resolution spectroscopy (HRS) technique is rooted in one of
the most prolific means of exoplanet detection to date: the Doppler
method, also known as the radial velocity technique. In the standard
Doppler method, photons from the planet are not detected, thus the
planet's existence is inferred. The standard method uses very precise,
high-resolution, stable spectrographs to measure the small red- and
blue-shift of the host star's spectrum as it moves toward and away
from us in response to the gravitational pull of its exoplanet
throughout its orbit. The star's radial velocity along our
line-of-sight is typically at the 100 meter-per-second (m/s) level for
close-in hot Jupiter exoplanets, and approaches the 10
centimeter-per-second (cm/s) level for Earth-like planets orbiting
Sun-like stars at $\sim 1\,$AU. The radial velocity semi-amplitude
($K_{\star}$) of the host star also enables us to determine a lower
limit on the planet's mass ($M_{p}$) via:

\begin{equation}
K_{\star}=\left(\frac{2\pi G}{P}\right)^{1/3}\frac{M_{p}\sin{i}}{(M_{p}+M_{\star})^{2/3}}\frac{1}{(1-e^{2})^{1/2}}
\end{equation}

where $P$ is the planet's orbital period, $e$ is its orbital
eccentricity, and $i$ is its orbital inclination, such that
$i=90^{\circ}$ is edge-on and $i=0^{\circ}$ is face-on along our
line-of-sight (see \citealt{Lov10} for a full derivation of this
equation). It gives only a lower limit on $M_{p}$ as we cannot assume
that we see the full magnitude of the planet's velocity vector, on
account of the unknown orbital inclination. The standard Doppler
method delivered the first confirmed detection of an exoplanet
orbiting a main-sequence host star: 51 Peg b \citep{May95}, which is a
non-transiting planet \citep{Hen97,Walk06,Brog13,Bir17}. But it did
not give us the spectrum of the planet. To get this, we must instead
consider what is happening with the other partner in this Doppler
dance i.e. the motion of the planet itself. The gravitational pull of
the star imparts a significantly larger orbital velocity to the
planet, and so the planet has a radial velocity that is typically
orders of magnitude faster than the star, ranging from
$K_{\oplus}\sim30\,$km/s for the Earth-Sun system, up to hundreds of
km/s for the closest-in hot Jupiters. Consequently, the red- and
blue-shift of the planetary spectral features are much larger than
those of the star, whose stellar spectral features are essentially
stationary by comparison. Furthermore, when observing from the ground,
any contamination from the Earth's atmosphere, such as telluric water
lines, is also essentially stationary. This is illustrated in
Figure~\ref{fig:technique1}. Therefore, we can use the large
wavelength shifts of the planet's spectrum during its orbit to
disentangle it from its host star and from our own atmosphere. For
this, we must perform high-resolution time-series spectroscopy,
observing often enough to sufficiently sample the Doppler shift of the
planet, i.e. long enough to detect notable wavelength shifts in the
planet's spectrum. The latter is set by the resolution of the
spectrograph, with the planet spectrum typically shifting across $>10$
pixels of the detector during the time series (see,
e.g. \citealt{Sne10,Bir13}).

The concept of using high-resolution spectroscopy to study exoplanet
atmospheres in this way arose not long after the discovery of 51 Peg
b. For example, the high-resolution ($R=100,000$) infrared
spectrograph CRIRES at ESO's Very Large Telescope (VLT) was designed
with this purpose in mind \citep{Wie96,Wei00}, aimed at detecting the
direct thermal emission spectra of exoplanets. Others searched instead
for the reflected light from the hot Jupiter $\tau$ boo b with optical
high-resolution spectrographs \citep{Char98,Char99,Col99,Col04},
although only found upper limits. Many of the diagnostic properties of
high-resolution spectra of exoplanets were outlined by \citet{Bro01}
and \citet{Spa02}, and multiple attempts were made to detect thermally
emitted light from giant exoplanets at infrared wavelengths. These
included using high-resolution spectrographs such as NIRSPEC on Keck
II and Phoenix on Gemini South
(e.g. \citealt{Bro02,Dem05,Barnes07,Barnes07b}) but again only
provided upper limits. It was only when \citet{Sne10} used CRIRES to
observe the hot Jupiter HD 209458 b that the technique delivered its
first robust detection of a molecule (carbon monoxide) in an exoplanet
atmosphere. Although poor weather may have inhibited earlier attempts
with other instruments, the stability of CRIRES, in part delivered by
its use of adaptive optics (to keep maximal flux of the star centred
on the slit entrance of the spectrograph) and Nasmyth mounting (to
stop the instrument moving around), and its superior spectral
resolution, were undoubtedly instrumental to its success. Since then,
high-resolution spectroscopy has been used to study the atmospheric
composition of both transiting and non-transiting close-in giant
planets
\citep{Brog12,Rod12,Bir13,deK13,Rod13,Brog13,Loc14,Brog14,Schw15,Hoe15,Brog16,Pis16,Bir17,All17,Brog17,Pis17,Nug17,Brog18},
and has provided some upper limits on mini-Neptune and super-Earth
atmospheres \citep{Cross11,Est17}. Note that while the majority of
these studies focused on detecting the dayside thermal emission of the
planet, several of them also observe the planet during its transit,
thus highlighting that traditional transmission spectroscopy is also
possible with HRS.

\begin{figure}
\includegraphics[width=\textwidth]{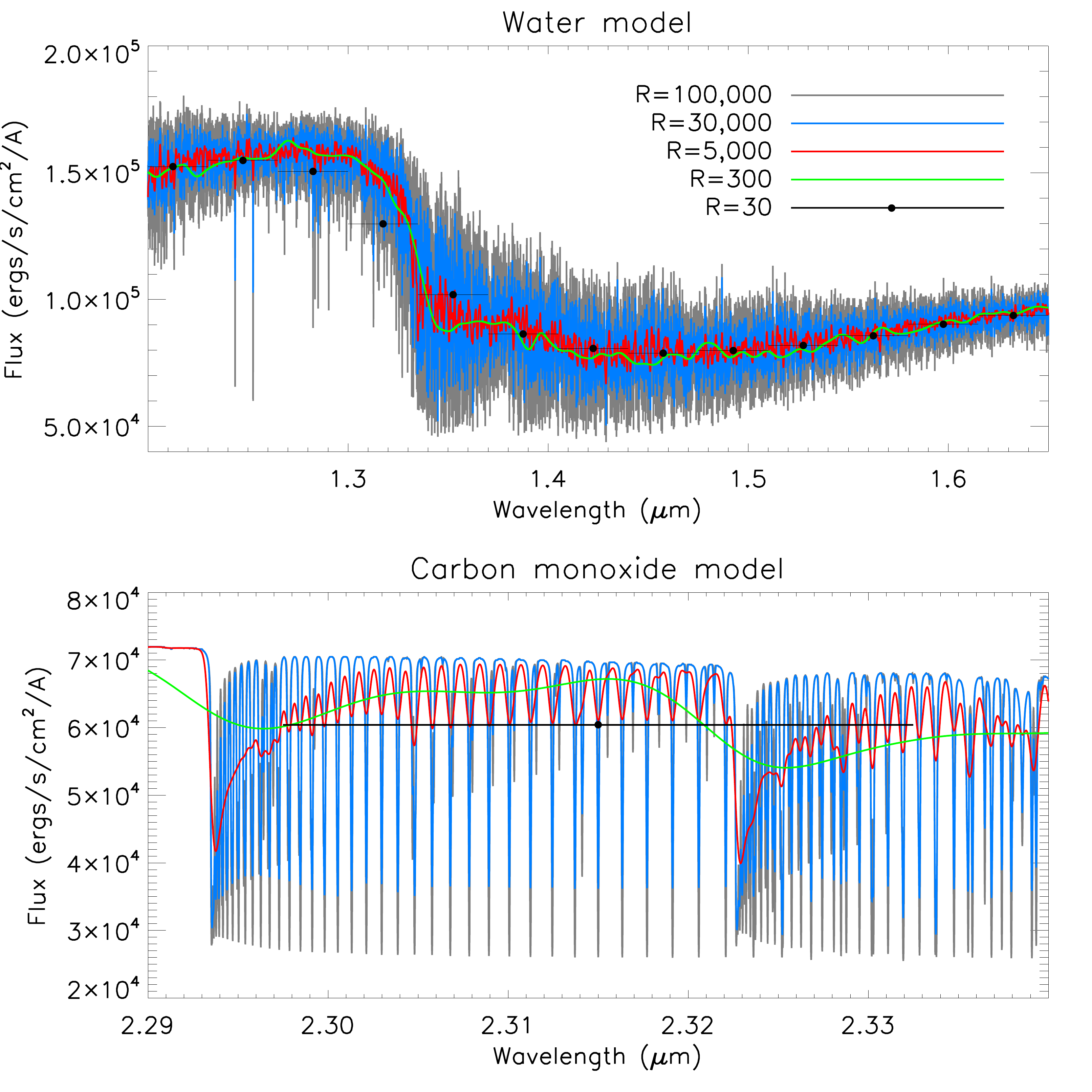}
\caption{The effect of decreasing spectral resolution. The two panels
  show different wavelength regions of a model hot Jupiter atmosphere
  containing water and carbon monoxide. Note the difference in the
  x-axis scale. The model has been convolved to different spectral
  resolutions. The overplotted points represent the typical resolution
  resulting from current space-based observations and trace out only
  broad molecular features. Note how many individual CO lines are lost
  between a resolution of R=100,000 and R=300. The shallower lines
  disappear more quickly, but some of the stronger CO lines remain
  even at R=5,000, albeit much reduced in line depth. Each line that
  is detected with the high-resolution technique increases the total
  planet signal-to-noise by a factor of $\sqrt(N_{lines})$.}
\label{fig:resolutions}
\end{figure}

Although we can use the Doppler shift to our advantage, there remains
the significant challenge of the planet-to-star flux contrast ratio,
which ranges from $10^{-3}$ to $10^{-10}$ for hot Jupiters to
Earth-Sun analogues, respectively, with the planet spectrum buried in
the noise of the stellar spectrum. This is where the increased
spectral resolution aspect of the HRS technique becomes crucial. Each
spectral line of the planet spectrum is detectable to a certain
significance, and the deeper the line, the easier it is to detect. The
more lines we can include in our measurement, the higher the
significance of our detection. At low spectral resolution, spectral
lines are convolved to shallower depths due to the limited resolving
power of the instrument, and also blended with other nearby
lines. Figure~\ref{fig:resolutions} demonstrates this effect. The
overplotted points in the figure represent the typical resolution
currently achieved with observations of exoplanets from e.g. the
Hubble Space Telescope \citep{Kre14b}, while the highest resolution
(R=100,000) represents ground-based spectrographs, such as
CRIRES. Space-based spectrographs are typically lower in resolution
because the increased volume, and hence weight of the instrument at
higher spectral resolution makes them too expensive and cumbersome to
launch. At high spectral resolution we better retain both the depth
and plurality of the spectral lines. By detecting multiple spectral
lines, we boost the signal-to-noise of the planet by the square root
of the number of lines we detect ($\sqrt{N_{lines}}$). This makes
high-resolution spectrographs with \emph{wide instantaneous spectral
  coverage} (i.e. number of wavelengths observed in a single exposure)
highly suited to the technique. Figure~\ref{fig:resolutions} also
highlights the very specific and \emph{unique} pattern of lines that
each molecule produces (water and carbon monoxide in a hot Jupiter are
shown as an example). These patterns are difficult to mimic by chance
at high spectral resolution, adding confidence to the detection of the
molecule. At lower resolution, great care must be taken to accurately
separate the blended, broad molecular bands of different molecules,
and to precisely remove any systematic effects that can introduce
mimics that are orders of magnitude larger than the planet
signal. Kreidberg's chapter of this handbook explains the removal of
systematics from lower-resolution spectra in detail.

Let us assess the signal-to-noise we can expect to achieve for a
planet using the HRS technique. For a close-in exoplanet, assuming
that we can completely remove the contaminating effects of the Earth's
atmosphere, we can write to first order the signal-to-noise ratio of
the planet spectrum obtained via HRS as follows \citep{Sne15}:

\begin{equation}
\rm SNR_{planet}=\frac{S_{p}\sqrt{N_{lines}}}{\sqrt{S_{\star}+\sigma_{bg}^{2}+\sigma_{read}^{2}+\sigma_{dark}^{2}}}
\end{equation}

where $S_{p}$ and $S_{\star}$ are the signal from the planet and star
respectively, both in units of photons per resolution
element. $\sigma_{bg}$, $\sigma_{read}$, and $\sigma_{dark}$,
represent noise from the sky and telescope background, the read-out
noise, and the dark current from the detector,
respectively. $N_{lines}$ is the number of lines \emph{detected} in
the observed wavelength range, i.e. it accounts for both the
\emph{plurality} and \emph{strength} of the lines in the planet
spectrum, and is thus dependent on the spectral resolution. In the
photon-limited regime, this reduces to the equation below:

\begin{equation}
\rm SNR_{planet}=\left(\frac{S_{p}}{S_{\star}}\right) SNR_{star} \sqrt{N_{lines}}
\label{eqn:SNR}
\end{equation}
where $\rm\left(\frac{S_{p}}{S_{\star}}\right)$ is the planet-to-star
signal ratio (e.g. the planet-star flux ratio or the amplitude of the
expected transmission signal) and $\rm SNR_{star}=\sqrt{S_{\star}}$ is
the photon-limited, \emph{total} signal-to-noise of the host star. For
example, assuming blackbody radiation, let's take a hot Jupiter and
host star with a dayside flux contrast ratio of
$5\times10^{-4}$. Let's also assume our spectrograph throughput
enables a signal-to-noise ratio on the star of 200 in 3 minute
exposure, and that we observe continuously for a half night ($\sim5$
h, $\sim100$ exposures), targeting the CO molecule. Let's also assume our
spectrograph has sufficient resolution to resolve 30 lines in the
$2.3\,\mu$m CO ro-vibrational overtone. The total
$\rm SNR_{star}\sim2000$ results in
$\rm SNR_{p}=5\times10^{-4}*200*\sqrt{100}*\sqrt{30}\sim$5.5. This
assumes that all the detected CO lines are equally and maximally deep.

\subsubsection{Seeing through the Earth's atmosphere}\label{sec:tellurics}
To detect the planet's spectral lines, we must first remove the
contaminating telluric lines present in ground-based observations,
taking advantage of the fact they do not vary in wavelength over the
time span of the observations, unlike the planet. Before beginning
this step, it can be beneficial to perform the following
procedures. First, clean any remaining bad pixels from the extracted
spectra, then align them to a common wavelength grid. The alignment
removes drift caused by e.g. changes in the temperature of the
instrument or telescope (e.g. \citealt{Brog13}). To preserve the
behaviour of any systematic effects in the observations, it is also
more suitable to work with the spectra in their observed pixel spacing
rather than re-linearize them to regular wavelength grid. It is also
therefore advantageous to have a highly stable spectrograph and
consistent observing strategy that negates significant realignment of
the spectra, even at the sub-pixel level.

Finally, it can be very useful in the beginning to inject a model
planet spectrum into the data at an appropriate orbital velocity, to
act as a guide during the telluric removal process (see panel E of
Figure 3). The injected model should reasonably match the expected
atmospheric profile and composition, and should be only just detectable
so not to significantly alter the noise properties of the data. This
enables a check on when the telluric removal process begins to also
remove planet signal, and highlights if the telluric cleaning process
is altering the planet spectral lines. It also enables a means to
assign weighting over the spectral range of the dataset to account
for, e.g. noisy detectors that naturally give lower signal strength in
the template-matching process. If the model is poorly recovered in a
spectral order, then the real planet signal will likely also have
lower detectability in this range and can be down-weighted for the
template matching. Of course, the injected signal should be removed in
the final analysis of the spectra.

\begin{figure}
\centering
\includegraphics[width=\textwidth]{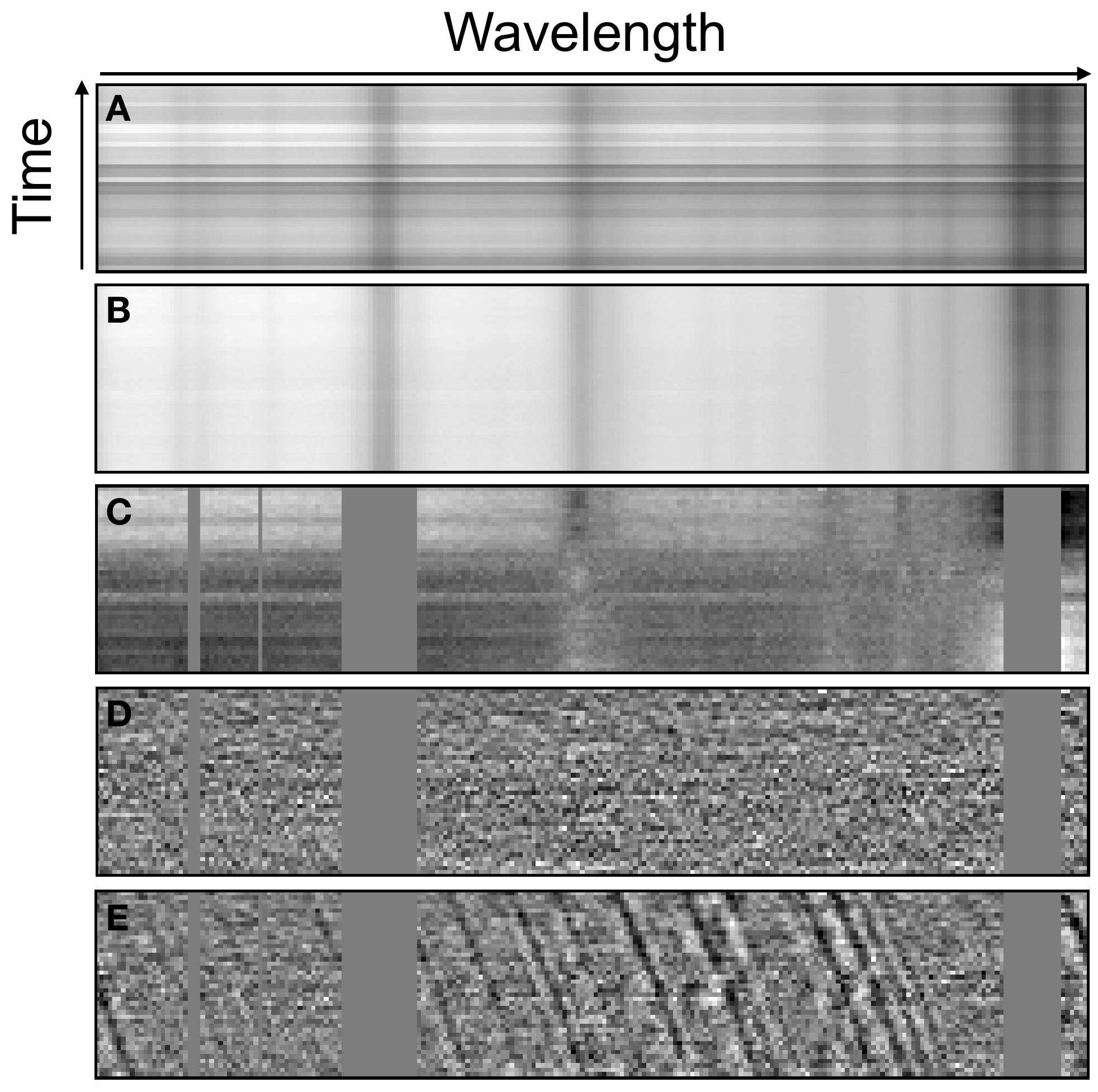}
\caption{Stages of the telluric removal process using principle
  component analysis. These data are from \citet{Bir17} who observed
  the dayside of 51 Peg b at $3.2\,\mu$m for 5 hours with
  CRIRES/VLT. There are 42 spectra in the time series. \textbf{A)} The
  observed spectra extracted from the CRIRES pipeline, after residual
  bad pixel cleaning and alignment to a common wavelength
  grid. \textbf{B)} The normalized spectra. \textbf{C)} After removing
  the first principle component. \textbf{D)} After removing optimal
  components, determined as when the planet signal starts to also be
  removed. The individual spectral lines of the planet are buried in
  the residual noise. \textbf{E)} The same as D but with a model water
  spectrum injected at stage A at $100\times$ the nominal planet
  single strength to highlight the positions of the planet's spectral
  lines. Solid gray regions are saturated telluric or stellar lines
  that have been masked. Note the gradual removal of the telluric
  feature in the center of each panel. The individual lines of the
  planet are blue-shifting as it moves toward us after superior
  conjunction, as shown by the slanted features in panel E, which
  trace out a small portion of the planet's RV curve. }
\label{fig:waterfall}
\end{figure}

Standard techniques to remove telluric lines typically use an observed
telluric standard A-star before or after each science spectrum to
correct the data (e.g. \citealt{Cross11}). However, because we want to
observe the planet lines Doppler shifting and gain as much
signal-to-noise as possible, we cannot continuously interrupt the
observation to take telluric standards. Sparsely observed telluric
standards are not a good representation of the atmosphere during the
spectral time series, and can therefore leave significant
residuals. Since the telluric lines only change in depth and not
position over time, an alternative approach is to simply model the
flux observed in each pixel of the spectrum over the time series. The
main cause of the variation in the telluric lines is usually changes
in air mass, followed by changes in e.g. telescope temperature,
in-slit position, and seeing. After normalizing the spectra, we can
use linear regression to form a function that describes the behaviour
of the tellurics in each pixel of the spectrum over time and remove
them (this is a column-by-column operation on the matrices shown in
Figure~\ref{fig:waterfall}). The final step is to divide each column
by its variance to restore the noise properties of the renormalized
data, and stop very noisy residual pixels from dominating. Any
remaining large scale, low-order gradients can be removed with a
high-pass filter applied to each spectrum. This has proven effective
in multiple planet detections
(e.g. \citealt{Sne10,Brog12,Brog13,Brog14,Brog17,Brog18}). In some
cases however, additional unknown systematics persist. In these cases,
we can instead adopt a blind approach to modelling the telluric
variations in each pixel over time. Singular value decomposition (SVD;
\citealt{Kal96}), or principle component analysis (PCA;
\citealt{Murt87,Pre92}), can be used to find common modes over time
for each pixel in the spectral time series. The idea here is that the
Doppler-shift of the planet moves its spectrum across the pixels
during the time series; thus it is not identified as a common mode for
any one particular pixel. The procedure removes the tellurics and
stellar continuum, leaving behind the continuum-normalized planet
spectrum buried in the residual photon noise. A demonstration of the
different stages of telluric removal with the PCA method is shown in
Figure~\ref{fig:waterfall}. Various implementations of this approach
have made multiple successful planet detections
(e.g. \citealt{deK13,Bir13,Pis16,Bir17,Pis17}), including algorithms
that enable weighting of the data by their errors, such as the popular
SYSREM algorithm typically applied transit survey light curves
\citep{Tam05}. However, in all of these approaches, the spectra are
normalized, which removes information about the true continuum level
of the planet. The removal of the continuum means that models with
different abundances and atmospheric temperature profiles can look
very similar, especially when comparing the depths of different lines
in the spectrum, and thus introduces degeneracies. An alternative
approach to remove the telluric features is to model them using
theoretical calculations based on the atmospheric conditions during
the exposure, using codes such as ESO's \textsc{Molecfit} or
\textsc{TERRASPEC} (e.g. \citealt{Rod12, Loc14, Sme15}). Modelling the
tellurics directly has the advantage of preserving the continuum of
the planet spectrum. On the other hand, it does not automatically
account for additional instrumental systematic effects occurring
during the observation, again leaving residuals. The best approach to
telluric removal is still under refinement, although the different
approaches can already produce residual spectra with noise approaching
their theoretical photon limit within $5-20\%$ \citep{Brog14}. Only in
the cores of the deepest telluric lines does the noise still
significantly exceed the photon limit (by a factor of a few), but
since these represent a small fraction of the data, the analysis of
HRS data overall approaches the theoretical photon limit, which
conversely is seldom achieved at lower spectral resolution.

In some cases, at certain wavelengths, the star may also contain
sufficiently similar spectral lines to the planet (e.g. CO lines at
$2.3\mu\,$m in hot Jupiters and cooler host stars such as K-dwarfs),
or may undergo non-negligible velocity changes (e.g. high mass ratio
systems such as a hot Jupiter orbiting a cool small M-dwarf), such
that it can interfere with the telluric removal process. The stellar
lines can be modelled and removed before tackling the
tellurics. \citet{Brog16} and \citet{Schw16} describe iterative
approaches to this with a model atmosphere that can also include
astrophysical alterations to the line profile, e.g.  the
Rossiter-McLaughlin effect during the transit of the planet.

\subsubsection{Finding the planet in the noise}
With the telluric and stellar lines removed from the spectra, we are
left simply with the residuals, as shown in the lower panels of
Figure~\ref{fig:waterfall}, which contain noise and the planet's
spectral lines. The individual planet lines have low signal-to-noise
($\rm SNR_{line}\sim<1$), so we must combine their signals to detect
the planet spectrum (as indicated in Equation~\ref{eqn:SNR}). A
powerful approach to this is the cross-correlation method, which uses
a template spectrum appropriate for the planet atmosphere to scan
through the residuals of each observation and compare how well the
template matches the specific line positions, and the ratio of the
line depths, in the observed planet spectrum. It is a fingerprint
matching exercise for exoplanet atmospheres.

The templates should match the spectral resolution and velocity
resolution of the data, and are generated by a comprehensive radiative
transfer treatment that models the interaction of light with the
matter in the atmosphere. Heng's chapter of this handbook introduces
this, and is elaborated in, e.g. \citet{Heng16}. The depth of the
lines and their relative depth ratios in the template spectra are
governed largely by the composition of the model atmosphere and its
temperature-pressure ($T-P$) profile. The latter describes how the
temperature changes in the atmosphere as a function of pressure (or
altitude). On Earth, the $T-P$ profile is partially inverted creating
a stratosphere i.e. there is a layer that heats up after an initial
cool-off with increasing altitude. It is caused by ozone absorbing UV
radiation and heating the atmosphere. $T-P$ inversions result in
emission lines, rather than absorption lines in the spectrum. A weak
inversion may only fill in the cores of the absorption lines, leaving
an apparently shallower feature suggesting an isothermal structure
instead (e.g. \citealt{Schw16,Parm18}). The higher the spectral
resolution, the easier it is to identify these more complicated line
shapes in the spectrum of the exoplanet.

Another crucial input to creating the template spectra is a precise
list of line positions and opacities for different molecules at
different temperatures and pressures. These are either measured in the
laboratory, or (in cases where it is not possible to do this safely)
with ab initio calculations from quantum chemistry. There are several
databases that provide such line lists and opacities, including HITRAN
and HITEMP \citep{Roth13,Gord17,Roth10}, and ExoMol \citep{Ten16}. At
low spectral resolution, inaccuracies in the individual line positions
are more tolerable as they are blended and convolved by the
instrument, as shown in Figure~\ref{fig:resolutions}. However, at high
resolution accurate line positions become crucial. Even if we have
used the correct properties for the planet atmosphere, the
cross-correlation will not get a good match if the theoretical or
laboratory-measured values of the line position are incorrect. We will
have lost our $\sqrt{N_{lines}}$ advantage. For simple molecules,
e.g. CO, the line positions are very well determined. However, for
more complex molecules such as water (H$_{2}$O) and methane
(CH$_{4}$), there remain some inaccuracies in the line positions at
high spectral resolution at temperatures appropriate for hot Jupiters,
as demonstrated in \citet{Harg15}. It highlights the importance of
laboratory research in the pursuit of understanding the full diversity
of the exoplanet zoo (e.g. \citealt{Hors18}).

\begin{figure}
\centering
\includegraphics[width=\textwidth]{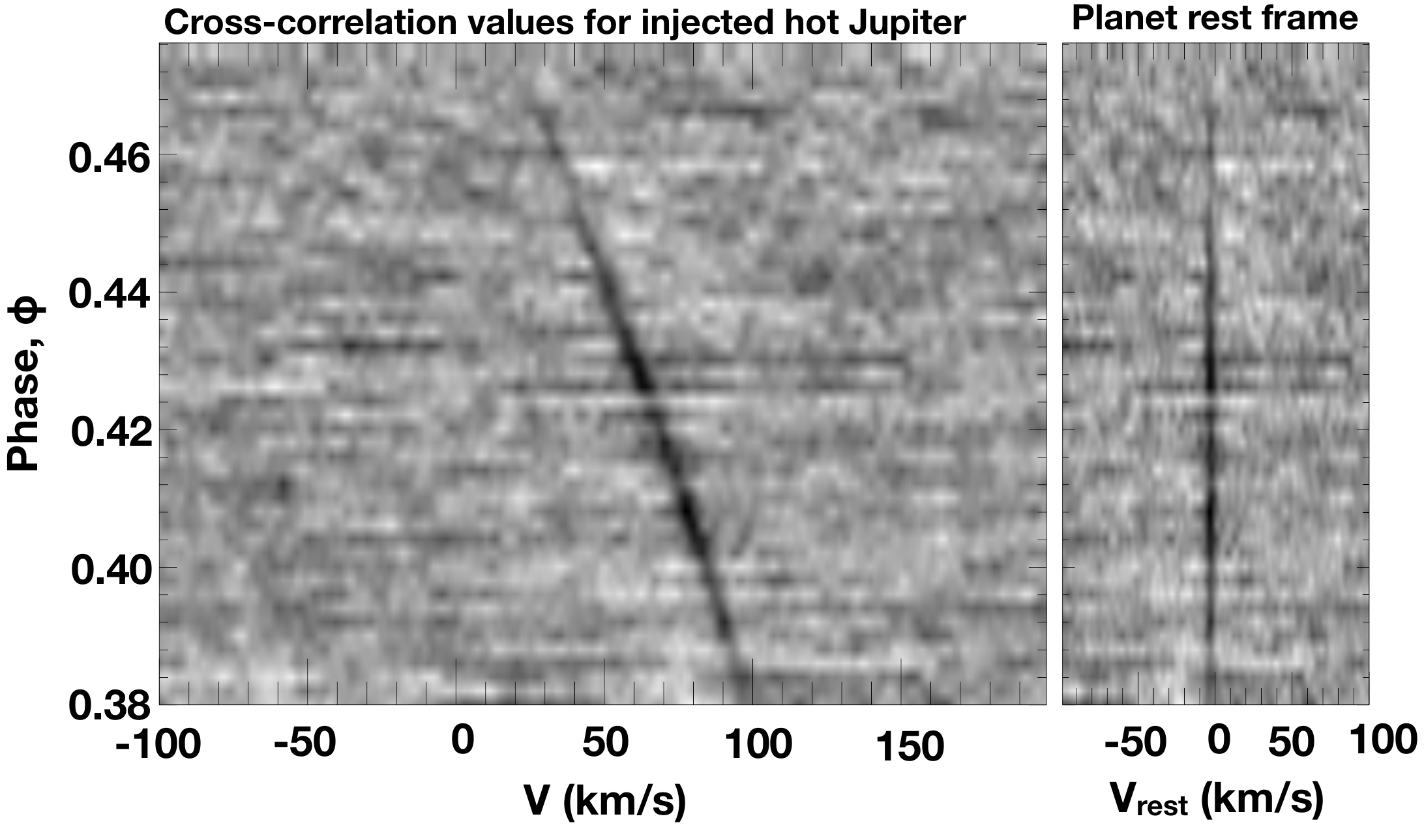}
\caption{Left: Part of the radial velocity curve observed over 5 hours
  for an injected model hot Jupiter. The RV trail is traced out by the
  blue-shifting peak of the cross-correlation functions (CCFs). The
  model contains water lines and was injected into real observations
  of HD 189733 from \citet{Bir13} at $5\times$ the nominal model
  strength for clarity. The slope of the trail corresponds to the RV
  semi-amplitude of the planet, $K_{p}$. Right: The CCFs shifted into
  the planet rest frame, where $K_{p}=154\,$km/s. The aligned CCFs can
  be summed into a single CCF for each trial $K_{p}$, which creates
  each row in Figure~\ref{fig:matrix}.}
\label{fig:xcor}
\end{figure}

Once we have generated a suite of templates that could plausibly match
the planet's observed spectrum, we can use them to perform
cross-correlations to find the planet signal. Each cross-correlation
gives a detection strength (i.e. how well the template matches the
data) for each trial wavelength shift (or velocity shift). The step
size between trial shifts should be no smaller than the velocity
resolution of the pixels on the detector to avoid oversampling the
data. Figure~\ref{fig:xcor} highlights how the peak of each
cross-correlation function (CCF) moves in response to the Doppler
shift of the planet, tracing out part of its radial velocity curve
from which we can determine the radial velocity semi-amplitude of the
planet, $K_{p}$. For transiting planets, we already know this velocity
because $\sin{i}$ is solved. For non-transiting planets, we must find
$K_{p}$ by determining the slope of the CCF trail. To do this, we can
shift the CCFs into the rest frame of the planet, assuming a given
$K_{p}$. The total Doppler shift of the planet not only includes its
radial velocity ($V_{RV}$, which can account for eccentric orbits),
but also the center-of-mass velocity of the star-planet system with
respect to the Earth ($V_{sys}$). This is typically measured using the
standard Doppler method on the host star. Finally, we must include the
velocity of the observer i.e. the velocity induced by the motion of
the Earth around the Sun (the barycentric correction, $V_{bary}$). We
can write the total planet velocity ($V_{p}$) as:

\begin{equation}
V_{p}=V_{RV} + V_{sys} + V_{bary}
\label{eqn:velocity}
\end{equation}

\begin{figure}
\centering
\includegraphics[width=\textwidth]{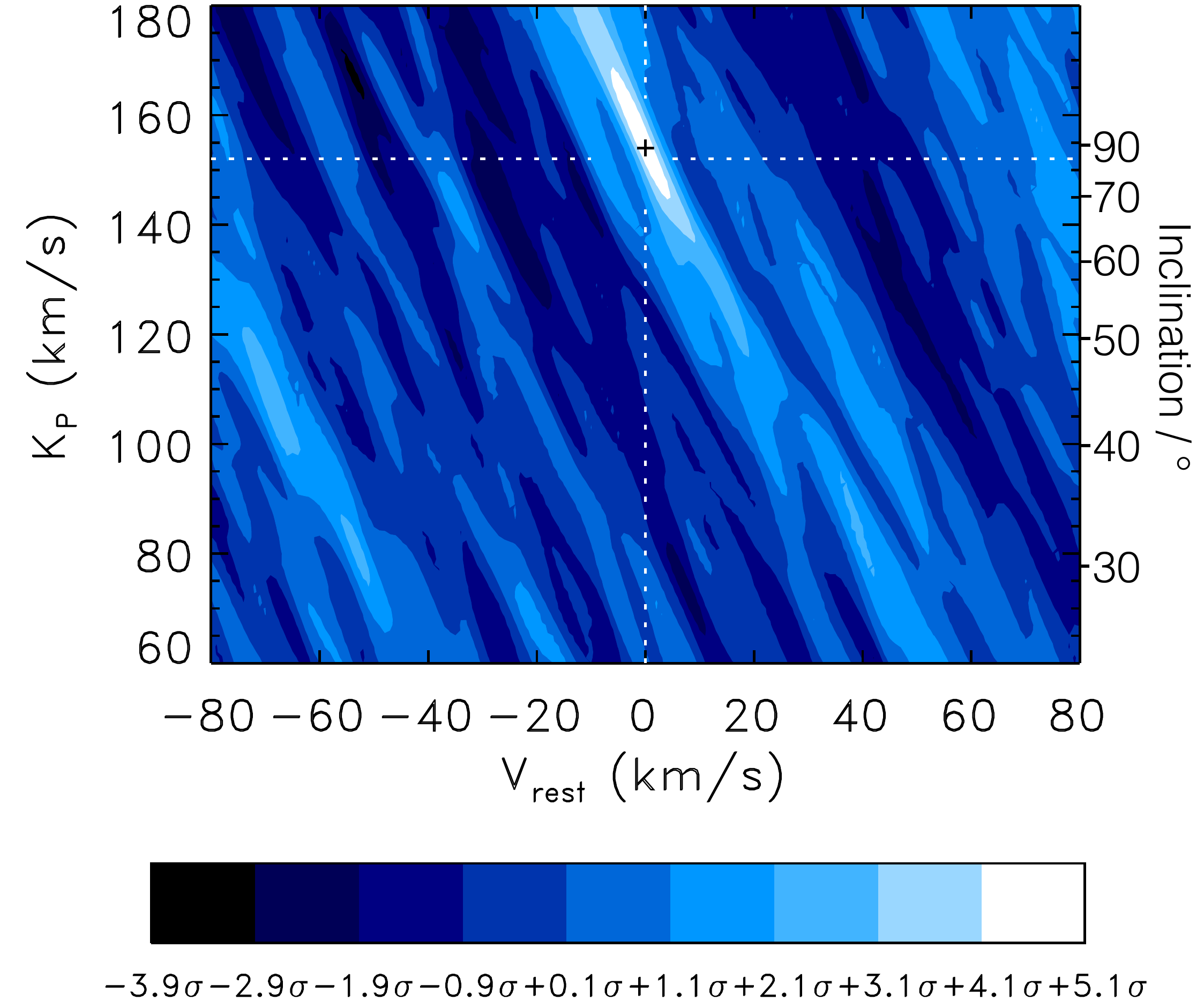}
\caption{A $5\sigma$ detection of water in the hot Jupiter HD 189733 b
  using the high-resolution spectroscopy technique. Each row in the
  matrix is a summed cross-correlation function for a given planet RV
  semi-amplitude ($K_{p}$). When the correct $K_{p}$ value is
  approached, the individual peaks of the CCFs align and contribute
  maximally to the signal. The white contour represents the $1\sigma$
  error ellipse in $K_{p}$ and $V_{sys}$, and is coincident with the
  rest frame velocity of the planet (dotted lines). This velocity
  includes the center-of-mass velocity of the star-planet system, and
  helps confirm the planetary nature of the signal. Note that there
  are regions corresponding to $3\sigma$ significances, highlighting
  the importance of exploring the parameter space around the expected
  $V_{sys}$ and $K_{p}$ of the planet, even in transiting cases. The
  planet mass and its orbital inclination can be derived given the
  value of $K_{p}$, as shown in Equations~\ref{eqn:mass}
  and~\ref{eqn:inclination}. The derived inclinations for given
  $K_{p}$ values are shown on the right-hand y-axis.}
\label{fig:matrix}
\end{figure}

Once in the planet rest frame, we can sum the CCFs, combining signal
from all of the lines in all of the observed spectra. If we repeat
this process for a range of different $K_{p}$, we find which $K_{p}$
gives the strongest peak in the summed CCF. This is shown in
Figure~\ref{fig:matrix}. As we approach the true $K_{p}$ of the planet
and the CCF peaks align, we see a stronger peak in the summed
CCF. Importantly, this peak should occur at zero velocity in the
planet rest frame as it confirms that the signal is at the same
$V_{sys}$ as the host star. Significant deviations from $V_{sys}$
generally indicate that the signal is not planetary in
nature. However, in some cases, physical phenomena occurring in the
planet's atmosphere, such as strong day-to-night winds and heat
circulation can cause offsets of a few km/s
(e.g. \citealt{Sne10,Kemp12, Show13,Brog16,Zha17}). This is discussed
further below.

The final step in securing the exoplanet's spectroscopic detection is
to determine the significance level of the signal. The distribution of
the cross-correlation values is Gaussian (see, e.g. \citealt{Brog12}),
so we can make a simple approximation of the significance of the
planet signal by dividing the CCFs by their standard deviation. A more
robust approach is to compare the CCF values in the radial velocity
trail of the planet to those outside it, and use a Welch T-test
\citep{Wel47} to determine the probability that both populations are
drawn from the same parent distribution. This can be done at each
trial $K_{p}$. Figure~\ref{fig:matrix} shows the distribution of these
probabilities. Values of $\leq3\sigma$ can be seen throughout the
parameter space and are false positives, or aliases generated by the
autocorrelation function of the template. It is therefore important to
explore the region around the expected planet rest frame velocity and
expected $K_{p}$ to determine the errors on the measured
quantities. We could go back to the residual spectra and align and sum
them into the planet rest frame using the measured $K_{p}$ to create
an actual spectrum of the planet. However, given that each line has
very low signal-to-noise, the spectral features are difficult to
discern. For example, for a planet detected at $6\sigma$ using 50
spectra with each containing 30 strong CO lines, then the individual
lines in the summed spectrum would be expected to be present at a
significance of $\sim1.1\sigma$ and at $<0.2\sigma$ per
spectrum. However, in some cases, it is possible to see the strongest
lines of, e.g. carbon monoxide, extending clearly from the noise (see,
e.g. \citealt{Brog12,Schw16}).

An alternative approach to detecting the planet spectrum is to leave
the stellar spectrum intact and treat the system in a similar way as
stellar binaries, using techniques such as Todcor \citep{Zuc94}. This
performs a cross-correlation with two templates simultaneously, one
for the star and one for the planet. It works optimally when the lines
of the planet and stellar spectra are maximally separated (i.e. at
quadrature), unlike the previously described method which works best
when the planet is undergoing its greatest rate of velocity change
(i.e. before and after superior/inferior conjunction, but avoiding
exact conjunction times due to the star and planet spectral lines
overlapping). Todcor yields a two-dimensional array of CCFs for each
component of the system per spectrum. These can be be combined into
nightly stellar and planetary likelihood curves in which to
search for peaks caused by the planet at $V_{sys}$ (see examples in
\citealt{Loc14,Pis16,Pis17}). This approach can be more beneficial in
systems that undergo less velocity change e.g. giant planets on longer
orbital periods, and helps probe the exoplanet population that move
too slowly for HRS, but are still too close to their host star to
isolate, even with the assistance of high-contrast imaging (as
described further below).

\subsubsection{Model-independent planet masses and radii}
The measurement of $K_{p}$ converts the star-planet system into a
double-lined spectroscopic binary. We can therefore derive the true
mass of non-transiting planets, despite not knowing the orbital
inclination, via the velocity and mass ratios:

\begin{equation}
\frac{K_{\star}}{K_{p}}=\frac{M_{p}}{M_{\star}}
\label{eqn:mass}
\end{equation}

where $M_{\star}$ is determined from stellar evolution models. We can
then even derive the planet's orbital inclination in non-transiting
systems, via:

\begin{equation}
\sin{i}=K_{p}/V_{orb}
\label{eqn:inclination}
\end{equation}

assuming a circular orbit, where $V_{orb}$=$2\pi a/P$, and that the
mass of the planet is negligible compared to the host star mass. The
typical precision in the derived planet mass is at the $5\%$ level
\citep{deK13} and acts as proof of the planetary mass nature of
non-transiting systems; a particularly important aspect for our
nearest rocky neighbours. If the planet is also a transiting system,
HRS converts it to a double-lined eclipsing binary system, allowing
model-independent measurements of the mass and radius of the host star
and its planet. Such systems are important calibrators for stellar
evolution models, especially for M-dwarfs, the most common stars in
our local neighbourhood. Stellar models tend to underpredict the radii
of small stars by $\sim3-15\%$ (see
e.g. \citealt{Lop05,Tor10,Irw11,Bir12,Bar15,Ditt17,Lub17,Jack18,Kes18}),
which can lead to a miscalculation of the planet's radius and thus its
atmospheric properties.

\subsection{Widely-separated exoplanets}
Exoplanets on wide orbital paths move slowly. This means their
spectral features do not change their Doppler shift significantly
during the course of a nightly observation, and thus require a
different approach to isolating their spectra from the glare of their
host stars and Earth's telluric features. Traditional high-contrast
imaging (HCI) techniques suppress a fraction of the light of the star
at the planet's position, thus improving the star-planet contrast
ratio. An example of this is adaptive optics (AO), which corrects for
the turbulence in the Earth's atmosphere and aims to deliver
diffraction-limited images. This not only creates a very sharp image
of the star and a better spatial resolution of the star and planet but
also adds stability to the observations, ensuring that the star does
not wander around on the detector (or in the entrance of a
spectrograph). We saw in the previous sections that high-resolution
spectroscopy further enables us to increase the signal-to-noise of the
planet by combining many of its spectral lines via
cross-correlation. The combination of these two techniques (HRS+HCI)
is a powerful tool for exoplanet spectroscopic detection and
characterization, and is described in detail by \citet{Spa02} and
\citet{Sne15}. In the case of HRS+HCI, the Doppler-shift advantage is
replaced by the fact that the planet's spectrum is uniquely different
to the star and \emph{strongly localized}. In a simple long-slit
spectrograph, the star fills most of the slit and its light is
detectable over many rows on the detector, while the fainter planet
spectrum is typically only detectable in one row. This requires the
entrance slit of the spectrograph to be aligned to contain both the
star and the planet. In an integral field unit (IFU), every ``pixel''
(or spaxel) is effectively an entrance to the spectrograph and thus
generates its own spectrum. We can therefore use rows or spaxels
containing only the stellar spectrum and tellurics as a template that
can be scaled and removed from every row or spaxel, leaving behind
only the planet spectrum at its unique location, buried in the
residual noise. The planet is then detected by cross-correlating the
residuals in each row or spaxel with a model template, with the
strongest peak in the CCFs occurring at the planet's position. Note
that the planet's spectrum will still contain its radial velocity at
the time of observation, providing a data point on the planet's
long-period RV curve, which helps determine its orbit in combination
with astrometry from direct images (e.g. \citealt{Schw16}).

We can rewrite the equation for the expected signal-to-noise of the
planet using HRS+HCI as follows:

\begin{equation}
\rm SNR_{planet}=\left(\frac{S_{p}}{S_{\star}}\right) SNR_{star} \sqrt{c_{HCI}N_{lines}}
\end{equation}

where $c_{HCI}$ is the suppression factor of the stellar flux at the
planet position.  It is important to note that if $c_{HCI}$ is too
strong, we fall out of the photon-limited noise regime and into the
sky background-limited regime. Beyond $5\mu$m, sky background begins
to dominate; hence, in cases of strong $c_{HCI}$ traditional direct
imaging processing techniques may be more sensitive than HRS+HCI. The
first successful demonstration of HRS+HCI was achieved by
\citet{Sne14}, who observed the young giant exoplanet $\beta$ Pic with
CRIRES/VLT. The planet was detected at its known spatial location by
matching with a model atmosphere spectrum containing carbon monoxide,
and its measured RV showed a blue-shift that helped determine the
likelihood of it transiting its host star in 2017-2018. The planetary
mass companion GQ Lup b was also detected using this technique,
appearing at its known location matching with carbon monoxide and
water model atmosphere spectra, at a signal-to-noise of
$SNR_{p}\sim12$ and clear CO absorption lines in the extracted planet
spectrum \citep{Schw16}. Even at $R=4000$, cross-correlation can still
detect the presence of CO and water in the atmospheres of widely
separated planets \citep{Barm15}. \citet{Hoe18} further demonstrate
the enormous potential of using moderate-resolution IFUs to create
``molecule maps'' which locate the position of the planet and detect
its atmosphere by combining HRS+HCI, opening new avenues for direct
imaging survey strategies.

\subsection{Atmospheric properties}
The use of model atmospheric spectra in the high resolution
spectroscopy technique means that it not only detects the planet, but
simultaneously characterizes its atmosphere. By using model
atmospheres (rather than a binary mask), the technique goes beyond
revealing the mere presence of a molecule, and places constraints on
the chemical abundances in the atmosphere, as well as the
temperature-pressure profile. The more lines and molecules that HRS
can detect, the stronger its constraint on the planet's atmospheric
composition. To determine some the planet's atmospheric properties in
the cross-correlation analysis, it can be helpful to inject the
best-matching model into the observed spectra (at the planet velocity)
with different negative scaling factors, and re-run the full analysis
starting at the telluric removal. When the planet signal is completely
canceled (a $0\sigma$ detection), the line contrast can be measured
i.e. the depth of the deepest planet lines with respect to the
continuum, divided by the stellar flux. This line contrast, which is
convolved by the resolution of the spectrograph, indicates the
difference between the line cores in the planet spectrum and its
broadband continuum, and thus helps constrain the T-P profile and
abundances in the planet atmosphere. Multiple models from the grid of
T-P profiles and abundances tried in the cross-correlation could
provide matches within $1\sigma$ of the best-matching model, and thus
provide an estimate of the errors on these properties. \citet{deK13}
highlighted pathways to achieve tight constraints on abundances with
HRS alone, proposing spectrographs with a wide instantaneous spectral
coverage as key instruments for its success. They also note that an
additional advantage of high-resolution spectroscopy is its ability to
probe above cloud decks where the cores of the strongest spectral
lines are formed. Lower-resolution spectra see only muted spectral
lines; thus, HRS potentially enables measurements of the composition
of cloudy worlds that have eluded lower-resolution observations (see,
e.g. \citealt{Kre14a}).

\subsubsection{A multi-resolution approach}
\citet{deK14} highlight the power of combining both low-resolution
spectroscopy (LRS) and HRS measurements, which can tightly constrain
the $T-P$ profile and continuum level of the planet, thus removing
degeneracies with composition. \citet{Brog17} demonstrated this using
multi-resolution spectra of HD 209458 b, determining the planet's
metallicity to $0.1-1\times$ its host star value. \citet{Pin17} also
demonstrated the effectiveness of multi-resolution spectra by
confirming that the flat spectrum of HD 189733 b (likely due to
scattering from aerosols) seen at low resolution can be reconciled
with the sodium spectral feature seen at high
resolution. Multi-resolution spectroscopic observations are therefore
highly valuable in characterizing exoplanet atmospheres, and hence in
understanding their formation history and evolution.

\subsubsection{Spinning worlds}
A unique aspect of HRS is that it is sensitive to the precise shape of
the lines in the planet spectrum. For a rotating planet passing
through transit, the leading limb of the atmosphere is red-shifting
while the trailing limb is blue-shifting. However, the disk-integrated
observation of current HRS merges these into a single broadened
spectral line that results in a CCF for the planet that is broader
than expected from the model and instrument resolving power. This
effect was seen the case of the widely separated giant exoplanet
$\beta$ Pic b, resulting in the first measurement of the length of a
day on an exoplanet. Its rotational velocity along our line-of-sight
was $V_{rot}\sim25$ km/s, corresponding to a day length of $\sim8$
hours \citep{Sne14}. This is faster than Jupiter's $\sim10$ hour day,
and the young exoplanet is expected to further spin up as its cools
and contracts. GQ Lup b on the other hand only rotates at $\sim5$
km/s, which may be due to its young age \citep{Schw16}. Understanding
how planets spin and acquire their angular momentum is a key part of
their formation history. Planets that have migrated to very close-in
orbits are expected to be tidally locked to their host
star. \citet{Brog16} measured the rotational period of the hot Jupiter
HD 189733 b to be $P_{rot}=1.7^{+2.9}_{−0.4}$ days, which is
consistent within the errors with the orbital period of $\sim2.2$ days
and significantly slower than observed for younger exoplanets that are
not tidally locked.

\subsubsection{Windswept worlds}
Another advantage of direct spectroscopic detection with HRS is its
sensitivity to small velocity shifts arising from dynamical processes
in the exoplanet atmosphere. The original HRS detection of CO during
the transit of HD 209458 b was accompanied by a $2\pm1$ km/s
blue-shift in the planet spectrum in excess of its expected total
velocity (based on equation~\ref{eqn:velocity}). This was interpreted
as a fast wind blowing from the highly irradiated dayside of the
planet all around the limb to its nightside and was supported by
theoretical prediction \citep{Kemp12,Show13,Kemp14}. A similarly fast
wind was also observed later during the transit of the hot Jupiter HD
189733 b, detected at both optical and infrared wavelengths
\citep{Lou15,Brog16}, via independent analysis
techniques. \citet{Lou15} were able to further demonstrate, by
studying the line profile of sodium in the planet's optical
transmission spectrum, that the winds on this hot Jupiter form an
eastward equatorial jet, with leading and trailing limbs of the planet
giving different velocity offsets. Measuring wind speeds may help us
understand why hot Jupiters have larger radii than expected, by
revealing how they dissipate kinetic energy in the presence of
magnetic fields (see e.g. \citealt{Kol18}, and Laughlin's chapter of
this handbook for more discussion on this issue). Simulations also
predict that east- or westward hot spots on hot Jupiters could be
revealed by velocity excesses when observing the dayside of the planet
with HRS \citep{Zha17} and may allow a measurement of the planet's
magnetic field strength \citep{Rog17}.

\subsection{A high-resolution future for exoplanet atmospheres}
Despite the exciting advances in exoplanet detection and
characterization already achieved with high-resolution spectroscopy,
the technique is still in its infancy. This is largely due to the
scarcity of high-resolution spectrographs, with CRIRES/VLT (8-m) at
$R=100,000$ being the primary resource for the technique between 2010
and 2014, after which CRIRES was removed for upgrades. However,
\citet{Loc14} demonstrated that HRS would work at $R=25,000$ using
NIRSPEC/Keck II (10-m), trading spectral resolution for increased
telescope aperture and wider wavelength coverage. The instrument
variables of spectral resolution, throughput, aperture size, and
instantaneous wavelength coverage can therefore be balanced such that
smaller telescopes could be as powerful as the VLT for HRS for the
brightest giant exoplanets. \citet{Brog18} confirmed this by detecting
water vapour in HD 189733 b with GIANO ($R=50,000$) on the 3.6-m
TNG. The upgraded CRIRES+ returns to the VLT in 2019, with a wider
instantaneous spectral coverage that will enhance the signal-to-noise
of the planet detection via $\rm\sqrt{N_{lines}}$. Moreover, a large
suite of high-resolution spectrographs will soon come online across
the globe, driven by the radial velocity hunt for Earth-like
planets. This requires very stable, high-resolution spectrographs at
optical or infrared wavelengths making them also ideal instruments for
HRS of exoplanet atmospheres. Some of the forerunners include the
infrared spectrographs: iSHELL/IRTF, CARMENES/CAHA, GIANO/TNG,
SPIRou/CFHT, ARIES/MMT, IGRINS/Gemini, HPF/HET, IRD/Subaru,
iLocator/LBT, CRIRES+/VLT, and NIRSPEC/Keck, to name but a few, while
the optical will see great advances with e.g. ESPRESSO/VLT and
EXPRES/DCT (see \citealt{Wri17} for a comprehensive list). This suite
will enable the study of smaller, cooler planets with HRS than
achieved so far. Looking further ahead to the mid-2020s, the first
light instruments in the era of the extremely large telescopes (ELTs),
also include high-resolution spectrographs e.g. METIS/ELT
\citep{Bran14} in the infrared and G-CLEF/GMT \citep{Sze16} in the
optical. The ELTs will make substantial advances with HRS, including
the detection of less abundant chemical species in exoplanet
atmospheres. This is important for identifying biomarkers in rocky
planet atmospheres, as well as isotopologues that might indicate the
evolutionary history of the planet. It will also enable phase-resolved
(e.g. dusk/dawn, day/night) studies of exoplanet atmospheres.

One of the most tantalizing prospects for future instrumentation is
the detection of an atmosphere surrounding our nearest potentially
habitable neighbour, Proxima b \citep{Sne15,Lov17,Wan17}. The oxygen
band at 760nm is an ideal target for HRS+HCI in this non-transiting
system. The reflected light from Proxima b will be Doppler-shifted,
but it will not contain many photons from the thermal spectrum of the
planet. Instead it will be dominated by the stellar spectrum but
modulated by the planet's reflection spectrum, adding additional
components to the model templates needed when cross-correlating. The
development of HRS for detecting and interpreting reflected light from
exoplanets is under way (see
e.g. \citealt{Char99,Lei03,Col04,Mart15,Hoe17}). In the exceptional
case of a transiting, habitable zone Earth-like planet discovered
orbiting a nearby M-dwarf, the oxygen band is again accessible with
HRS during the planet's transit. The expected depth of the lines is
only $\sim3\times$ smaller than for CO in a hot Jupiter, but the host
star is likely to be very faint and thus requires the
photon-collecting power of an ELT \citep{Sne13,Rod13}.

Finally, for the widely separated planets, the ELTs may make it
possible to map out features such as large storms in their
atmospheres, due to the perturbations that they cause in the spectral line
profile. This is similar to how spots are mapped out on stars, or the
Rossiter-McLaughlin effect during a planetary transit. It uses Doppler
tomography, i.e. it studies how the line profile (or shape of the
cross-correlation function) changes as a feature rotates in and out of
view. Such a map was recently created for a nearby brown dwarf using
the VLT, giving promise to the technique's success to map out the
appearance of gas giant exoplanets in the future
\citep{Sne14,Cross14map}.

\begin{acknowledgement}
  JLB thanks Eleanor Spring, Matteo Brogi, Ignas Snellen, Henriette
  Schwarz, Jens Hoeijmakers, Ernst de Mooij, and Remco de Kok for
  helpful discussions on this chapter. This work was performed in part
  under contract with the Jet Propulsion Laboratory (JPL) funded by
  NASA through the Sagan Fellowship Program executed by the NASA
  Exoplanet Science Institute.
\end{acknowledgement}
\bibliographystyle{spbasicHBexo}
\bibliography{referencesjlb}
\end{document}